\documentclass[aps,prl,twocolumn,floatfix]{revtex4-2}
\usepackage[utf8]{inputenc}
\usepackage[english]{babel}
\usepackage{mathtools}
\usepackage{newtxtext}  
\usepackage{amsmath}
\usepackage{amsfonts}
\usepackage[varbb,varg]{newtxmath} 
\usepackage{bm,dsfont,eucal}
\AtBeginDocument{\renewcommand{\vec}[1]{\bm{#1}}}

\usepackage[dvipsnames,svgnames,x11names,hyperref]{xcolor}
\usepackage{hyperref}
\hypersetup{ 
    colorlinks=true, linkcolor=NavyBlue, urlcolor=NavyBlue, citecolor=NavyBlue
}
\usepackage{physics,braket,siunitx,slashed}
\usepackage[capitalise]{cleveref}

\begin{document}
        \title{Superpolarized Electron-Hole Liquid and Multiferroicity in Multilayer Graphene} 

\author{Mainak Das and Chunli Huang}
\affiliation{Department of Physics and Astronomy, University of Kentucky, Lexington, Kentucky 40506-0055, USA}

\date{\today}

\begin{abstract}We introduce a many-body state termed superpolarized electron-hole liquid to explain the multiferroic properties observed in a recent experiment on rhombohedral pentalayer graphene by Han et al. [\href{https://doi.org/10.1038/s41565-023-01520-1}{Nature 623, 41-47, 2023}] . Superpolarization refers to a state where electrons and holes are fully polarized in opposite directions within the extended spin-valley space, resulting in a polarization per charge that exceeds the saturated value of one. This state exhibits multiferroic order, characterized by spontaneous spin, layer, and valley polarization. We demonstrate the independent control of valley polarization and orbital magnetization in this state through the application of electric-displacement field and weak magnetic fields. The large magnetoelectric effect observed in the experiment is attributed to the substantial Berry curvature concentrated near the band edge. The concept of superpolarization can be probed experimentally through magnetic oscillation and local current distribution measurements to determine the area of the Fermi surfaces and their magnetic moments.
\end{abstract} 

\maketitle

\textit{Introduction:~}
The exploration of strongly correlated states in multilayer graphene \cite{de_la_Barrera_2022,doi:10.1126/science.abm8386,seiler2022quantum,zhou_half_2021,zhou_superconductivity_2021,lee2019gate,PhysRevB.80.035429,PhysRevLett.125.227702,PhysRevB.109.L060409,lee2014competition,liu2023spontaneous,Myhro_2018,PhysRevLett.120.096802,wirth2022experimental,PhysRevB.105.L201107,PhysRevB.82.035409,PhysRevB.107.L121405,PhysRevB.88.075408,chatterjee2022inter,das2023quarter,arp2023intervalley,PhysRevLett.122.046403,zhang_band_2010,koshino_trigonal_2009} has been significantly advanced by recent experiments in rhombohedral pentalayer graphene (rPG) \cite{Han_2023,han2023orbital,lu2023fractional}. Within the density and electric displacement-field phase space, these experiments have identified fractional Chern insulators, integer Chern insulators with large Chern numbers, and a multiferroic metal exhibiting a significant magnetoelectric effect, whose microscopic nature remains elusive.  
In this letter, we demonstrate that the multiferroic metal in rPG can be understood as a novel magnetic state, which we term the superpolarized electron-hole liquid.

Low density electrons in multilayer graphene can be appropriately described as a generalized two-dimensional electron gas model where each electron possesses an octet pseudospin degree of freedom, consisting of layer, valley, and spin. These electrons interact with each other through repulsive long-range Coulomb interactions. The layer pseudospin degree of freedom pertains to the outermost layers ($A_1-B_N$), while the valley refers to the two inequivalent corners of the honeycomb Brillouin zone. A primary approximation for understanding magnetism in this system is the Stoner mechanism \cite{seiler2022quantum,zhou_half_2021,zhou_superconductivity_2021,lee2019gate,PhysRevB.80.035429,PhysRevB.109.L060409,dong2023stoner}, wherein these itinerant electrons collectively polarize in the octet space to minimize the (pseudospin-independent) repulsive Coulomb interaction. The direction of polarization depends on smaller energy scales such as spin-orbit coupling and lattice-scale coupling \cite{wei2024landau}. Recent experiments have made progress in determining the direction of this polarization and estimating the strength of intrinsic spin-orbit coupling \cite{arp2023intervalley} in the quarter metal region. A quarter metal, representing one of the simplest forms of generalized ferromagnetism in multilayer graphene at finite electric displacement field, spontaneously distributes all electrons or holes into one of the spin-valley flavors, resulting in finite order parameters $\braket{\hat{s}_z}$ and $\braket{\hat{\tau}_z}$. Since the layer polarization $\braket{\hat{\sigma}_z}$ is induced by an external electric displacement field $D$ in the experiment \cite{zhou_half_2021}, this spin-valley imbalanced quarter-metal also exhibits a non-vanishing orbital magnetization $M_{\text{orb}} \propto \braket{\hat{\sigma}_z\hat{\tau}_z}$. In Ref.~\cite{han2023orbital}, the number of graphene layers was increased to five and they found that the symmetry-broken metal persists even at vanishing $D$ at moderate doping ($|n_e| < 10^{12} \text{cm}^{-2}$). A central question arises: Is there finite spontaneous layer polarization at $D=0$, and what determines the direction? How does generalized ferromagnetism in terms of layer-spin-valley contribute to a $D$-dependent anomalous Hall effect  with a hysteresis that manifests an intriguing butterfly-shaped pattern?

We address all these questions using self-consistent Fock calculations, incorporating experimentally informed electronic structure models for rhombohedral graphene, the recently identified strength of intrinsic Ising spin-orbit coupling \cite{arp2023intervalley,das2023quarter}, and the dual-gated screened Coulomb interaction. These calculations yield multiple possible symmetry-breaking ground states, whose energy differences would naturally depend on the parameters of the Hamiltonian. However, stringent experimental data allows us to unambiguously identify the microscopic state that explains the multiferroic behavior as the superpolarized electron-hole liquid (SEHL). In what follows, we first discuss the band structure of SEHL, highlighting the concept of superpolarization and its ferroic orders. We then examine the magnetoelectric effect in SEHL and analyze the magnetic and electric hysteresis observed in experiments, showing how alternative candidate states fail to explain this behavior.

\textit{Superpolarized Electron-Hole Liquid}: 
The mean-field Hamiltonian we consider is given by the following:
\begin{align}
\hat{H}^{MF}_{\vec{k}}=\hat{T}_{\vec{k}}+\hat{\Sigma}^F_{\vec{k}}
\label{label:MF_Hamiltonian}
\end{align} 
Here, $\hat{T}_{\vec{k}}$ represents the 40 by 40 band Hamiltonian for Bloch waves at momentum $\vec{k}$, arising from five layers and two sublattices, and incorporating both spin and valley degrees of freedom. The Slonczewski-Weiss-McClure parameters that describe this Hamilonian is described in the supplementary material \cite{supmat}.
Here $\hat{\Sigma}^F_{\vec{k}}=-\int \frac{d^2k'}{(2\pi)^2} V_{\vec{k}-\vec{k}'} \hat{\rho}_{\vec{k}'}$ is the Fock self-energy accounts for dual-gate screened Coulomb interactions of Fourier component $V_{\vec{q}}=2\pi k_e\tanh{(|\vec{q}|d)}/(\epsilon_r|\vec{q}|)$. Here, $k_e=1.44$ eV nm is the Coulomb constant, $\epsilon_r=25$ is screening constant, $d=5$ nm is the distance from gate to the material,
and $\hat{\rho}_{\vec{k}}$ is the density matrix constructed from the eigenspectrum of self-consistent Hamiltonian. 
\begin{table}[t]
  \begin{center}
  \vline
    \begin{tabular}{l|c|c|c|c|}
    \hline 
       & $
\left(\nu_{K\uparrow},
\nu_{K\downarrow},\nu_{K'\uparrow} ,
\nu_{K'\downarrow} \right)$& 
    $\braket{\hat{s}_z}=\braket{\hat{\tau}_z}$ & $\langle \hat{\sigma}_z\rangle$&  $\langle \hat{\sigma}_z\hat{\tau}_z\rangle$
    \\
      \hline 
      \text{VIQM} & $(1,0_\pm,0_\pm,0_\pm)$ & $1$ & $\pm$ & $\pm$ \\
        \hline
      \text{PM} & $\left(0.25,0.25,0.25,0.25\right)$ & $0$ & 0 & 0\\
      \hline
      & \color{red}{$(\nu_e,0_-,0_-,\nu_h) $}  & \color{red}{$\nu_h-\nu_e>1$} & \color{red}{+}&\color{red}{0}  \\
      \cline{2-5}
      \text{SEHL} & $(\nu_e,0_+,0_-,\nu_h) $  & $\nu_h-\nu_e>1$ & 0&-\\
      \cline{2-5}
      & $(\nu_e,0_-,0_+,\nu_h) $  & $\nu_h-\nu_e>1$ & 0 &+ \\
      \cline{2-5}
      & $(\nu_e,0_+,0_+,\nu_h) $  & $\nu_h-\nu_e>1$ & - & 0  \\
      \cline{2-5}
      \hline
    \end{tabular}
  \end{center}
\caption{Metals observed at low density in the multilayer graphene electron gas. Here, $\nu_h > 0$ and $\nu_e < 0$ represent the fractional populations of holes and electrons, respectively, with their difference equaling the total hole density in the experiment, i.e., $\nu_h + \nu_e = 1$. The row highlighted in red is favored by spin-orbit coupling and exhibits spontaneous spin polarization $\braket{\hat{s}_z}$, valley polarization $\braket{\hat{\tau}_z}$, and layer polarization $\braket{\hat{\sigma}_z}$, but exhibits vanishing orbital magnetization  $M_{\text{orb}}\propto\braket{\hat{\sigma}_z\hat{\tau}_z}$.
}
\label{tab:states}
\end{table}

At zero-displacement field, as we move toward charge neutrality from a hole density of $n_e = -6 \times 10^{11}$ cm$^{-2}$, mean-field calculations reveal ground state transitions from the valley-Ising quarter metal (VIQM) to the superpolarized electron-hole liquid (SEHL), and finally to the paramagnetic (PM) metal \cite{supmat}.
These states can be differentiated by the relative population of carriers within each flavor, $f={K\uparrow, K\downarrow, K'\uparrow, K'\downarrow}$, 
\footnote{The valley-XY (or IVC) state does not constitute the ground state in this region of the parameter space}:
\begin{equation}
\vec{\nu} =
\left(\nu_{K\uparrow}
\nu_{K\downarrow},
\nu_{K'\uparrow},
\nu_{K'\downarrow}\right), \label{eq:relative_charge_dist}
\end{equation}
Here, $\nu_{f} \equiv \frac{n_{f}}{n_e}$ is the filling fraction of flavor $f$, with the sum of the four components of $\vec{\nu}$ equaling 1. When a flavor has no carriers and exhibits a band gap, we denote it as $0_{\pm}$, where the $\pm$ subscript indicates the sign of the mass. This notation helps differentiate various forms of metals observed in experiments, as shown in Table~\ref{tab:states} 
\footnote{We can distribute holes into any 4 flavors, then the electrons can occupy any one of the remaining 3 flavors. Once these flavors are occupied, there are additional four possibilities for choosing the sign of the (topological) mass-term in the remaining two flavors. This leads to a total of 48 degeneracies.}

We define ``superpolarization'' as a condition where the groundstate expectation value of a quantity, such as spin polarization $\langle \hat{s}_z \rangle = \frac{1}{N}\Tr(\hat{s}_z \hat{\rho})$, surpasses the  saturated level of $\langle \hat{s}_z \rangle = 1$. In this expression, $N = |n_e|A$, where $A$ is the sample area. Table~\ref{tab:states} shows many SEHL states, however, the row highlighted in red, representing a state characterized by spontaneous spin, valley, and layer polarization, yet exhibiting vanishing orbital magnetization (proportional to $\braket{\hat{\sigma}_z\hat{\tau}_z}$). This is the state favored by the small intrinsic spin-orbit coupling (SOC):
\begin{equation}
    \hat{H}^{SOC}=\lambda \hat{\sigma}_z\hat{\tau}_z\hat{s}_z
\end{equation}
Here $\lambda\sim50~\mu$eV is the strength of SOC estimated in our theoretical and numerous experimental studies \cite{das2023quarter,arp2023intervalley,PhysRevLett.122.046403}. Note the layer polarization operator $\hat{\sigma}_z\equiv \ket{B_5}\bra{B_5}-\ket{A_1}\bra{A_1}$.

The  bandstructure of this state is shown in Figure~\ref{fig:1}.
For the $K\downarrow,K'\uparrow$ projected bands, there is a large interaction-induced energy gap between the valence and conduction bands.
The itinerant carriers occupy the remaining two flavors, $K'\downarrow$ and $K\uparrow$, and they form a hole-like and electron-like Fermi surfaces. The hole fraction $\nu_{K'\downarrow}=\nu_{h}>0$ and the  electron fraction $\nu_{K\uparrow}=\nu_{e}<0$ accounts for all the carriers $\nu_h+\nu_e=1$. Their Fermi level is comparable to the  Coulomb-potential so there is no interaction induced gap. Unlike the valley-Ising quarter metal, which always exhibits finite orbital magnetization ($\braket{\hat{\sigma}_z\hat{\tau}_z} \neq 0$), the SEHL state achieves vanishing orbital magnetization by closing an energy gap through the redistribution of electrons into an insulating flavor, as we will now explain.

\begin{figure}[t]
    \centering
    \includegraphics[width=\columnwidth]{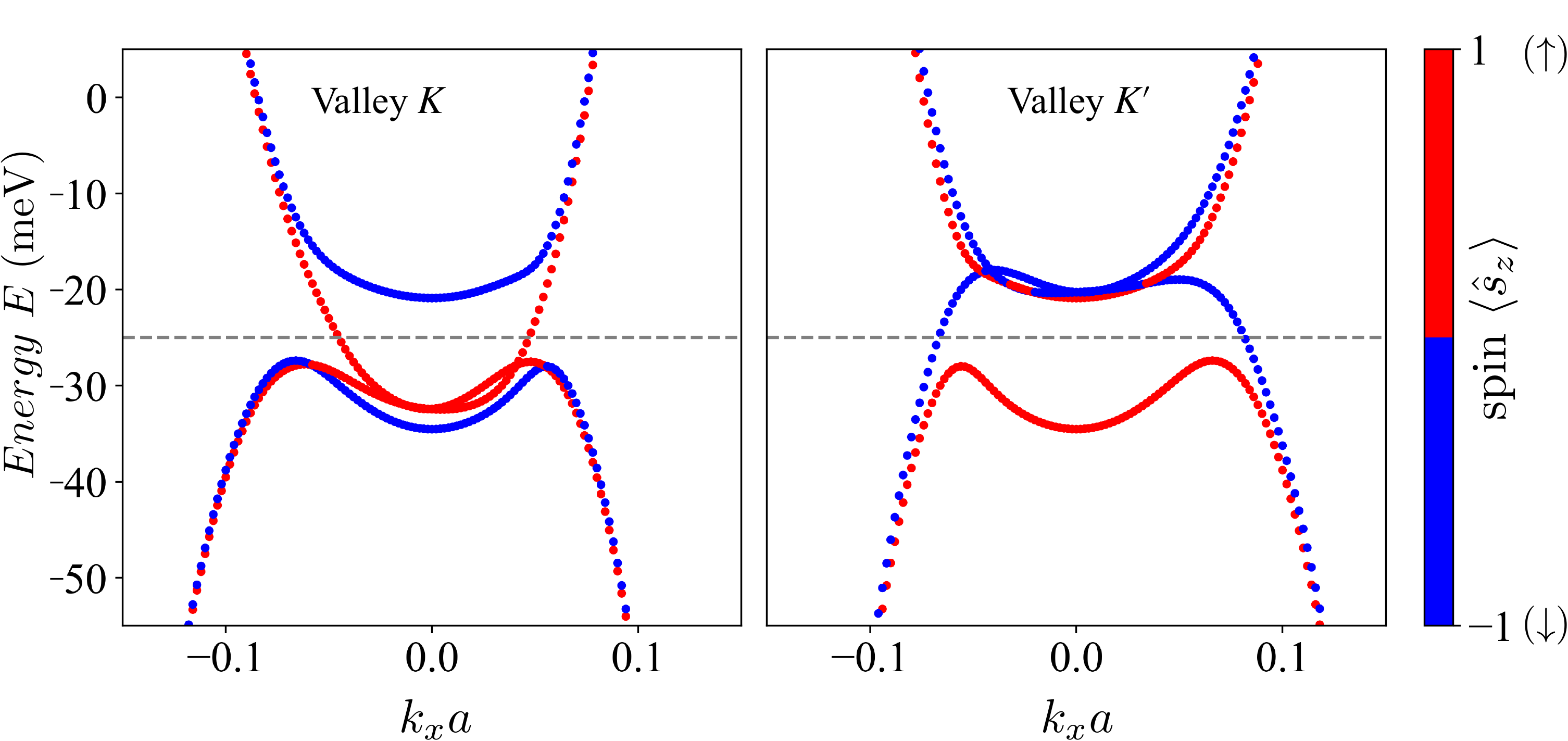}
    \caption{Bandstructure of superpolarized electron-hole liquid (SEHL) along $k_y=0$. The horizontal dashed line indicates the chemical potential $(\mu)$ at $(n_e,U)=-3\times 10^{11}$ cm$^{-2},0$ meV.}
    \label{fig:1}
\end{figure}

\textit{Orbital Magnetization and Magnetoelectric Effect:~}
\begin{figure*}[t]
    \centering
    \includegraphics[width=2\columnwidth]{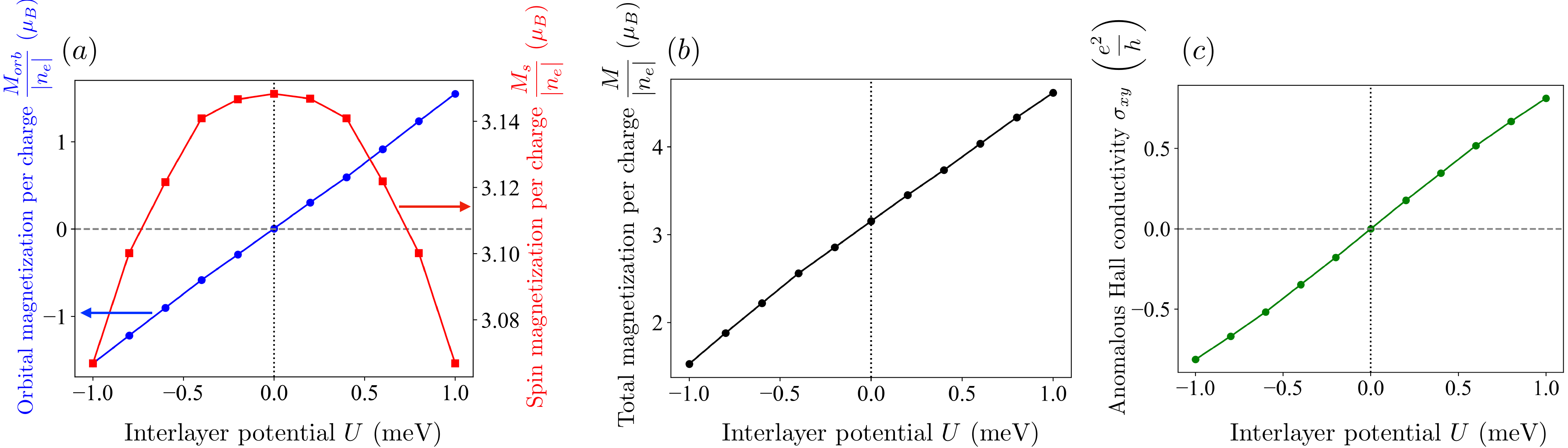}
    \caption{$(a)$ Induced magnetization due to interlayer potential $(U)$ on SEHL phase at $n_e=-3\times 10^{11}$ cm$^{-2}$ shows that spin magnetization is an even function of $U$, whereas orbital magnetization is an odd function of $U$. Also, $M^0_{\text{orb}}=0$, and $\alpha_{\text{orb}}\gg \alpha_s$. $(b)$ Total magnetization $(M)$ per charge vs. $U$ shows that $M$ can be expressed as $M\sim M_s^0+\alpha_{\text{orb}}U$. $(c)$ Anomalous hall conductivity $\sigma_{xy}$ changes linearly with $U$.}
    \label{fig:3}
\end{figure*}
The total magnetization $M$ can be decomposed into the spin contribution $M_s$ and the orbital contribution $M_{\text{orb}}$. Here, $M_s = \mu_B \int \frac{d^2k}{(2\pi)^2} \text{Tr}(\hat{\rho}_\mathbf{k} \hat{s}_z)$, while the orbital magnetization $M_{\text{orb}}$ is given by\cite{PhysRevB.109.L060409,shi2007quantum,PhysRevLett.125.227702}:
\begin{align}
M_{\text{orb}} &= \frac{e}{\hbar} \int \frac{d^2k}{(2\pi)^2}\sum_{n} n_F(\epsilon_{n,\vec{k}}) \sum_{n'\neq n}\left(\epsilon_{n,\vec{k}}+\epsilon_{n',\vec{k}}-2\mu \right)\notag\\ 
&\frac{\operatorname{Im}\left(\bra{\psi_{n\vec{k}}}\partial_x\hat{H}^{MF}_{\vec{k}}\ket{\psi_{n'\vec{k}}}\bra{\psi_{n'\vec{k}}}\partial_y\hat{H}^{MF}_{\vec{k}}\ket{\psi_{n\vec{k}}}\right)}{(\epsilon_{n,\vec{k}}-\epsilon_{n',\vec{k}})^2}
\end{align}
Here $\mu_B\sim 5.8\times 10^{-2}$ meV/T is Bohr magneton, $\epsilon_{n,\vec{k}}$ and $\ket{\psi_{n\vec{k}}}$ are the eigenspectrum of mean-Field Hamiltonian, $n_F(\epsilon)=1/(1+e^{\beta(\epsilon-\mu)})$ is Fermi-Dirac distribution, and $\mu$ is the chemical potential.
We study these magnetization contributions as a function of the applied interlayer potential $U$ in the SEHL phase. As shown in Fig.~\ref{fig:3}.(a), we observe that at $U=0$, the spin magnetization is finite, while the orbital magnetization vanishes. The orbital magnetization of the two insulating flavors $K\downarrow$ and $K'\uparrow$ sums to zero. The orbital magnetization of the other two flavors with electron-like and hole-like Fermi surfaces is independently zero. This is due to the $\mathcal{C}_2\mathcal{T}$ symmetry present in the valley-projected Hamiltonian. Here, $\mathcal{C}_2$ represents a $\pi$ rotation along the $z$-axis, and $\mathcal{T}$ denotes the time-reversal transformation. This symmetry ensures band touching between the valence and conduction bands, and the inter-band matrix elements of the velocity operator are real, $\bra{\psi_{n\vec{k}}}\frac{1}{\hbar}\nabla_{\vec{k}} \hat{H}^{MF}_{\vec{k}}\ket{\psi_{n'\vec{k}}} \in \mathbb{R}$. Consequently, the orbital magnetization, which is proportional to the imaginary part of the inter-band matrix element, vanishes unless the interlayer potential-induced gap breaks the $\mathcal{C}_2\mathcal{T}$ symmetry.
This is how the SEHL phase, a form of generalized ferromagnetism in multilayer graphene, exhibits spontaneous spin, valley, and layer polarization, yet the orbital magnetization vanishes.

However, when we examine the change in magnetization induced by the interlayer potential $U$, the orbital contribution significantly exceeds the spin contribution:
\begin{equation}
\frac{dM_{\text{orb}}}{dU} \gg \frac{dM_s}{dU}.
\end{equation}
This dominance of the orbital contribution is attributed to the significant concentration of Berry curvature at the band bottom in pentalayer graphene. Once a finite interlayer potential induces a band gap, the Berry curvature increases dramatically, leading to a pronounced response to the electric field. This effect is evident in both the orbital magnetization $M_{\text{orb}}$ and the anomalous Hall conductivity $\sigma_{xy}$. This is shown in Fig.\ref{fig:3}.(b) and (c).

The discussion above leads us to formulate a simple Ginzburg-Landau free-energy density $(F)$ that is useful for studying the magnetoelectric effect
\cite{Fiebig_2005,zvezdin2004phase,doi:10.1021/acsami.9b19320,PhysRevApplied.20.064011,PhysRevB.90.125412,he2020giant,cao2017moving,xie2022alternating}:
\begin{align}
F = E - M^0_{s} B -  \frac{|n_e|}{2}\braket{\hat{\sigma}_z}^0  U - \alpha_{\text{orb}}  U B 
\label{eq: Free_energy}
\end{align}
Here, $E$ 
is the ground state energy density, $M^0_s$ and $\braket{\hat{\sigma}_z}^0$ represent the spontaneous spin magnetization and spontaneous layer polarization, respectively. The coefficient $\alpha_{\text{orb}}=\partial M_{\text{orb}}/\partial U$, known as the orbital magnetoelectric susceptibility, a key indicator of the magnetoelectric effect. It characterizes the strength of layer polarization induced by a magnetic field, and vice versa. The unique property of multilayer graphene is that the spontaneous magnetization originates from spin magnetization ($M^0_s$), while the magnetoelectric susceptibility is dominated by the orbital contribution ($\alpha_{\text{orb}}$).

\textit{Hysteresis of Anomalous Hall Effect induced by electric and magnetic fields: }
\begin{figure*}[t]
    \centering
    \includegraphics[width=2\columnwidth]{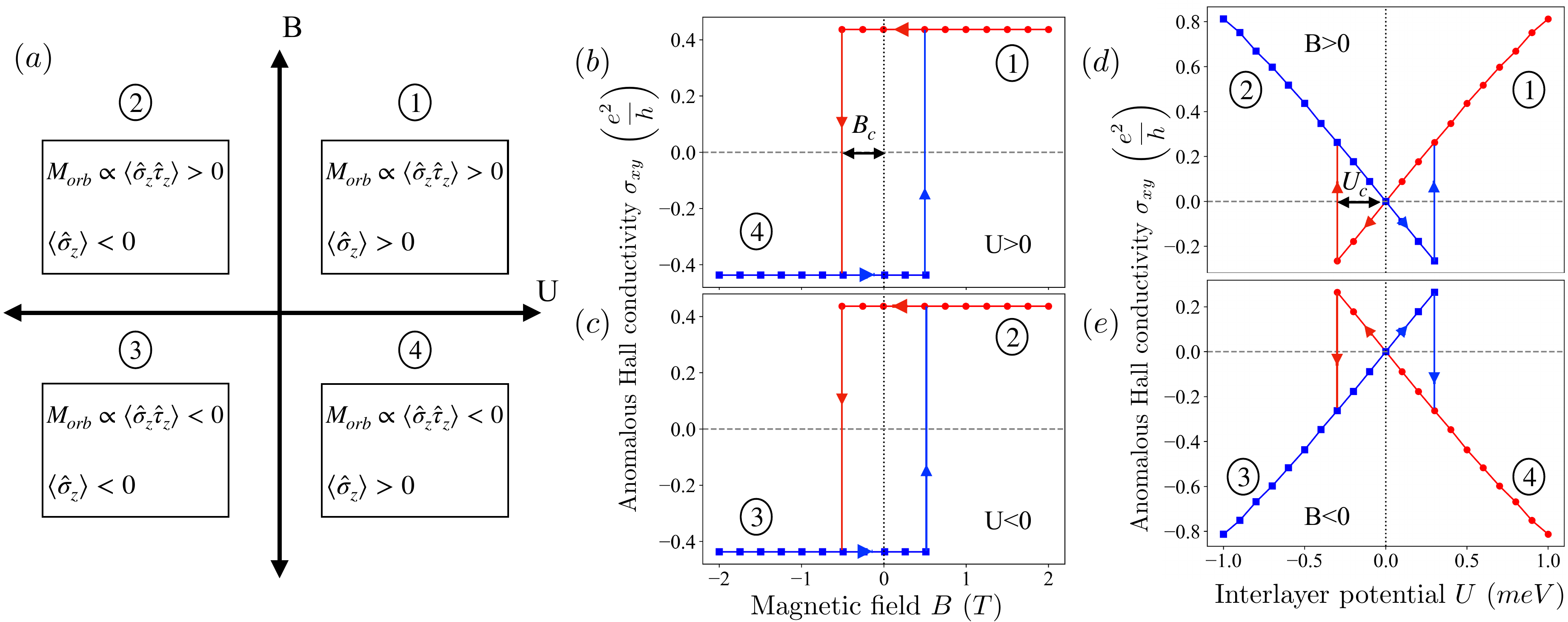}
    \caption{$(a)$ Interlayer potential $(U)$ and magnetic field $(B)$ select the groundstate (SEHL) orbital magnetization $(M_{\text{orb}})$ and layer polarization $\braket{\hat{\sigma}_z}$ independently. $(b)$ and $(c)$ Magnetic hysteresis at constant $U\lessgtr0$ suggests the chirality of the magnetic hysteresis is independent of $U$. $B_c$ indicates the magnetic coercive field. $(d)$ and $(e)$ Hysteresis in electric potential $U$ at constant $B$ shows butterfly-like pattern, and the electric hysteresis changes chirality under reversal of $B$-field. $U_c$ indicates the electric coercive potential. }
    \label{fig:4}
\end{figure*}
The order parameter of generalized ferromagnetism in multilayer graphene can exhibit large degeneracies in the octet layer-spin-valley space. For the superpolarized electron-hole liquid (SEHL), even in the presence of intrinsic spin-orbit coupling, there remains a fourfold discrete degeneracy. However, these degeneracies can be split using an external magnetic field $B$ and an electric potential $U$ that are perpendicular to the plane of the 2D material. Figure \ref{fig:4}.(a) illustrates that in different quadrants of the $U-B$ control parameter space, we can split the degenerate manifold with different hierarchies, and the lowest energy state is the one where the magnetization and layer polarization are aligned with the external fields. For example, for $B>0$, the ground state has spin magnetization $M_{s}>0$ and orbital magnetization $M_{\text{orb}}>0$, thus the order parameter $\langle \hat{s}_z\rangle > 0$ and $\langle \hat{\tau}_z \hat{\sigma}_z\rangle > 0$. When $U>0$, the layer polarization $\langle \hat{\sigma}_z\rangle > 0$. Therefore, in the first quadrant where $U>0$ and $B>0$, the ground state has $\langle \hat{s}_z\rangle > 0$, $\langle \hat{\sigma}_z\rangle > 0$, and $\langle \hat{\tau}_z \hat{\sigma}_z\rangle > 0$.

As we sweep the generalized ferromagnet through the $U-B$ control parameter space, the hierarchies of energy levels in the ground state manifold change, and their level crossings suggest a first-order phase transition. This transition can occur either by creating domain walls or through a single-domain reversal. The domain wall width can be estimated as the square root of the ratio of magnetic anisotropy energy to magnetic stiffness. When this width exceeds the system size, the phase transition is expected to be a single-domain. In this single-domain limit, we can estimate the coercive field using the spatially independent free energy described in Eq.~\eqref{eq: Free_energy}.

Let us now examine magnetic hysteresis at a constant interlayer potential (electric-displacement field). Since $U$ fixes the layer polarization ($\hat{\sigma}_z$), and $B\lessgtr0$ determines the direction of orbital magnetization ($M_{\text{orb}}\propto \langle \hat{\sigma}_z \hat{\tau}_z\rangle \lessgtr0$), the two energy levels that cross as we sweep between $\pm B$ at fixed $U$ have opposite valley polarization. This leads to the hysteresis in anomalous Hall conductivity, as shown in Fig.\ref{fig:4}b) and c).
The magnetic coercive field $B_c$ is estimated by equating the free energy difference of two SEHL phases with opposite magnetization to the magnetic anisotropy energy $E_{\text{aniso}}$:
\begin{align}
    2 (M_s^0+\alpha_{\text{orb}} U)\cdot B_c=  E_{\text{aniso}}.
    \label{eq: magnetic_coercive}
\end{align}We define the magnetic anisotropy energy $E_{\text{aniso}}$ as the energy difference between the valley-Ising state and the valley-XY (intervalley coherent state). Although the valley-XY state is a stable self-consistent Fock solution, its energy is higher than that of the simple valley-Ising state. The competition between these states is discussed in our earlier paper \cite{das2023quarter}. For pentalayer graphene, $E_{\text{aniso}}$ is approximately $0.2$ meV per charge. Note the coercive field $B_c$ decreases as electric displacement field increases. This effect is attributed to the magnetoelectric effect, where the electric displacement field increases the orbital magnetization by removing the band-touching between the valence and conduction bands. This electric field dependence of the $B_c$ is indeed observed in the experiment. We have reproduced this $B_c$ vs. $U$ in the supplementary material \cite{supmat}. Figure~\ref{fig:4}.(b) and (c) show that the chirality of the magnetic hysteresis curve, as we sweep back and forth between $\pm B$, is independent of the sign of the electric field.

Let us now explore the intriguing electric-field hysteresis of the anomalous Hall effect  at a constant applied magnetic field, that is, sweeping in the horizontal direction in Fig.~\ref{fig:4}.(a). Since the constant magnetic field $B$ fixes the orbital magnetization $\langle \hat{\sigma}_z \hat{\tau}_z\rangle$, and $U\lessgtr0$ determines the direction of layer polarization $\langle \hat{\sigma}_z \rangle\lessgtr0$, the two energy levels that cross as we sweep between $\pm U$ at fixed $B$ exhibit opposite senses of layer polarization and valley polarization. This simultaneous reversal of ground state layer polarization and valley polarization ensures that the orbital magnetization, and thus the anomalous Hall effect, is always determined by the right-hand rule of the applied external field. Consequently, the thermodynamically stable state always has the same sign of anomalous Hall conductivity $\sigma_{xy}$, as shown in Fig.\ref{fig:4}(d) and (e). This effect, combined with the fact that when $U=0$, the $\mathcal{C}_2\mathcal{T}$ symmetry ensures that $\sigma_{xy}=0$, gives rise to the butterfly-shaped hysteresis observed in the experiment.

Assuming single-domain reversal, we can estimate the critical electric coercive potential $U_c$ needed to reverse the ground state using Eq.~\eqref{eq: Free_energy}. We equate the free energy difference between states with opposite senses of layer polarization and valley polarization in the presence of a magnetic field to a magnetic anisotropy energy $E_{\text{aniso}}'$:
\begin{align}
2 (|n_e|\braket{\hat{\sigma}_z}^0/2 + \alpha_{\text{orb}} B) \cdot U_c = E_{\text{aniso}}'
\label{eq: electric_coercive}
\end{align}
We estimate $E_{\text{aniso}}'$ as the energy difference between the SEHL state and the layer unpolarized paramagnetic state and the value we found is around  $1$ meV per charge. Note that the critical electric coercive potential $U_c$ decreases as the magnetic field $B$ increases because the magnetic field can enhance layer polarization through the magnetoelectric effect ($\alpha_{\text{orb}} \neq 0$). The hysteresis curve of $\sigma_{xy}$ versus $U$ exhibits opposite chirality when we change the sign of $B$, as shown in Fig.~\ref{fig:4}(d) and (e). This reversal occurs because changing the sign of the magnetic field $B$ at constant $U$ reverses the ground state valley polarization $\langle \hat{\tau}_z \rangle$, without altering the layer polarization $\langle \hat{\sigma}_z \rangle$. Consequently, the anomalous Hall conductivity $\sigma_{xy} \propto \langle \hat{\sigma}_z \hat{\tau}z \rangle$, reverses sign.

We would like to point out that the simple valley-Ising quarter metal (VIQM), characterized by a single hole-like Fermi surface in one of the flavors, cannot account for the butterfly-shaped hysteresis observed in the experiment. While it is true that the change in magnetization induced by the electric field in the VIQM phase is also dominated by orbital magnetization \cite{PhysRevB.109.L060409}, the orbital magnetization $M_{\text{orb}}$ in VIQM peaks at $U=0$ and decreases with increasing $|U|$. 
Therefore, the hysteresis of $\sigma_{xy}$ in the VIQM phase would exhibit a diamond shape rather than a butterfly shape.

\textit{Discussion and outlook:} In summary, we have identified a novel magnetic metal in rhombohedral pentalayer graphene termed the superpolarized electron-hole liquid. This state is distinct from other magnetic metals such as the valley-Ising quarter metal  as it exhibits superpolarization, where spin and valley polarization exceed the conventional saturated level of one, due to the coexistence of electron and hole Fermi surfaces in different spin-valley flavors.
This phase enables independent control of orbital magnetization and valley polarization through external electric and magnetic fields, attributed to the redistribution of itinerant charges across distinct spin-valley flavors. The concept of superpolarization can be probed through a combination of magnetic oscillation and local current distribution experiments to determine the area of Fermi surfaces and their magnetic moments.

\textit{Acknowledgments:}  
We acknowledge useful discussions with Tonghang Han and Long Ju. We are grateful to the University of Kentucky Center for Computational Sciences and Information Technology Services Research Computing for their support and use of the Morgan Compute Cluster and associated research computing resources.

\bibliography{references}

\newpage
\widetext
\begin{center}
\textbf{\large Supplementary Materials:\\ Superpolarized Electron-Hole Liquid and Multiferroicity in Multilayer Graphene}
\end{center}
\setcounter{equation}{0}
\setcounter{figure}{0}
\setcounter{table}{0}
\setcounter{page}{1}
\makeatletter
\renewcommand{\theequation}{S\arabic{equation}}
\renewcommand{\thefigure}{S\arabic{figure}}

\maketitle
This supplementary material comprises five sections.
In the first section, we detail the construction of the non-interacting band Hamiltonian of rhombohedral pentalayer graphene (rPG). In the next section, we delve into the mean-field phase diagram of rPG in density $(n_e)$-interlayer potential $(U)$ space at the hole doping side $(n_e<0)$. Additionally, we illustrate the energy crossing between the valley-Ising quarter metal (VIQM) and the superpolarized electron-hole liquid (SEHL) at $U=0$.
The third section demonstrates the control of magnetic coercive field through external electric field, and vice versa.
In the fourth section, we project the layer polarization in to SEHL energy bands at $U=0$.
Finally, in the last section, we discuss the degeneracy in the SEHL phase in the absence of spin-orbit coupling (SOC) and any external field.

\section{A: Non-interacting band Hamiltonian}

In this section, we discuss the properties of the band Hamiltonian $\hat{T}_{\vec{k}}$ of rhombohedral pentalayer graphene (rPG) in the maintext. rPG contains five layers of carbon atoms. The interlayer distance is $0.34$ nm and each layer has two sublattices label by $\sigma=A,B$ with a lattice constant $a=0.246$ nm. The layers of rPG are in a ABCAB stacking configuration.
Following Refs.~\cite{zhang_band_2010,koshino_trigonal_2009}, we consider a single electronic orbital ($\pi$) per atom, with spin $s=\pm 1/2$, and use the continuum model that can be derived from a tight-binding Hamiltonian by expanding the low-energy dispersion around each valley $\tau$ with $\tau=\pm 1$ at the corners of the Brillouin zone. We label the four flavor combinations of spin $s$ and valley $\tau$ with the flavor index $f\equiv (\tau, s)$. In the basis $\ket{\psi_{f}(\vec{k})}=\left(\psi_{f A_1}(\vec{k}), \psi_{f B_1}(\vec{k}), \psi_{f A_2}(\vec{k}), \psi_{f B_2}(\vec{k}), \psi_{f A_3}(\vec{k
}), \psi_{f B_3}(\vec{k}), \psi_{f A_4}(\vec{k}), \psi_{f B_4}(\vec{k}), \psi_{f A_5}(\vec{k
}), \psi_{f B_5}(\vec{k})\right)$, the band Hamiltonian is given by
\begin{equation}
\hat{T}_{\vec{k}}=\sum_{s=\pm1,\tau=\pm1} \ket{\psi_{\tau,s}(\vec{k})} h_\tau(\vec{k}) \bra{\psi_{\tau,s}(\vec{k})}.
\end{equation}
The continuum Hamiltonian acting on the orbital space is different in opposite valley but spin-degenerate
\begin{align}\label{eq:continuum_hamiltonian}
    h_\tau(\vec{k}) = \begin{bmatrix}
    t(\vec{k}) + U_1 & t_{12}(\vec{k}) & t_{13}&0_{2\times 2}&0_{2\times 2} \\
    t_{12}^\dagger(\vec{k}) & t(\vec{k}) + U_2 & t_{12}(\vec{k}) &t_{13}&0_{2\times 2}\\
    t_{13}^\dagger & t_{12}^\dagger(\vec{k}) & t(\vec{k}) + U_3& t_{12}(\vec{k}) &t_{13}\\
    0_{2\times 2} &t_{13}^\dagger & t_{12}^\dagger(\vec{k}) & t(\vec{k}) + U_4& t_{12}(\vec{k})\\
    0_{2\times 2} & 0_{2\times 2}& t^\dagger_{13}& t_{12}^\dagger(\vec{k}) & t(\vec{k}) + U_5
    \end{bmatrix}_{10\times 10}.
\end{align}
Here $\vec{k}=(k_x,k_y)$ is the Bloch momentum measured with respect to the center of valley $\tau$. This Hamiltonian contains matrices with intra-layer hopping amplitudes ($t$), nearest-layer hopping amplitudes ($t_{12}$), and next-nearest-layer hopping amplitudes ($t_{13}$), as well as possible potential differences between different layers and/or sublattices due to external gates and/or broken symmetries (terms $U_i$). Explicitly, the various hopping amplitudes are given by
\begin{align}
    t(\vec{k}) = \begin{bmatrix}
      0 & v_0 \pi^\dagger \\
      v_0 \pi & 0
    \end{bmatrix}, \quad 
    t_{12}(\vec{k}) = \begin{bmatrix}
      -v_4 \pi^\dagger & v_3 \pi \\
      \gamma_1 & -v_4 \pi^\dagger
    \end{bmatrix}, \quad
    t_{13} = \begin{bmatrix}
      0 & \gamma_2/2 \\
      0 & 0
    \end{bmatrix}
\end{align}
where $\pi = \tau k_x+i k_y$ is a linear momentum, and $\gamma_i$ ($i=0,\dots, 4$) are hopping amplitudes of the tight-binding model with corresponding velocity parameters $v_i=(\sqrt{3}/2)a\gamma_i/\hbar$. The ${U_i}$ terms are
\begin{align}
    U_1 = \begin{bmatrix}
      \frac{\Delta+\delta+U}{2} & 0 \\
      0 & \frac{\Delta+U}{2}
    \end{bmatrix}, \quad 
    U_2 = \begin{bmatrix}
      \frac{-\Delta+\delta}{2}+\frac{U}{4} & 0 \\
      0 & -\frac{\Delta}{2}+\frac{U}{4}
    \end{bmatrix}, \quad
    U_3= \begin{bmatrix}
      \frac{\delta}{2} & 0 \\
      0 & \frac{\delta}{2}
    \end{bmatrix}, \quad 
    U_4 = \begin{bmatrix}
      \frac{-\Delta}{2}-\frac{U}{4} & 0 \\
      0 & \frac{-\Delta+\delta}{2}-\frac{U}{4}
    \end{bmatrix}, \quad 
    U_5 = \begin{bmatrix}
      \frac{\Delta}{2}-\frac{U}{2} & 0 \\
      0 & \frac{\Delta+\delta}{2}-\frac{U}{2}
    \end{bmatrix}
\end{align}
We use the model parameters listed in Table I, which are chosen to match quantum oscillation frequency signatures in Ref.~\cite{zhou_half_2021}.

\begin{table}[t]
\caption{\label{tab:graphene_params_ABC}Tight-binding parameters (in eV) for rhombohedral trilayer graphene, see also Refs.~\cite{zhang_band_2010,koshino_trigonal_2009,zhou_half_2021}.}
\begin{ruledtabular}
\begin{tabular}{llllllll}
$\gamma_0$ & $\gamma_1$ & $\gamma_2$ & $\gamma_3$ & $\gamma_4$ & $U$ & $\Delta$ & $\delta$ \\
$3.160$ & $0.380$ & $-0.015$ & $-0.290$ & $0.141$ & $0.030$ & $-0.0023$ & $-0.0105$ \\
\end{tabular} 
\end{ruledtabular} 
\end{table}
In the next section, we discuss the mean field phase diagram in rPG for hole doping $(n_e<0)$.

\section{B: Mean-field phase diagram} In this section, we discuss the mean-field phase diagram in $n_e-U$ space, and the nature of phase boundaries at hole doping side $(n_e<0)$.

\begin{figure}[h]
    \centering
    \includegraphics[width=\columnwidth]{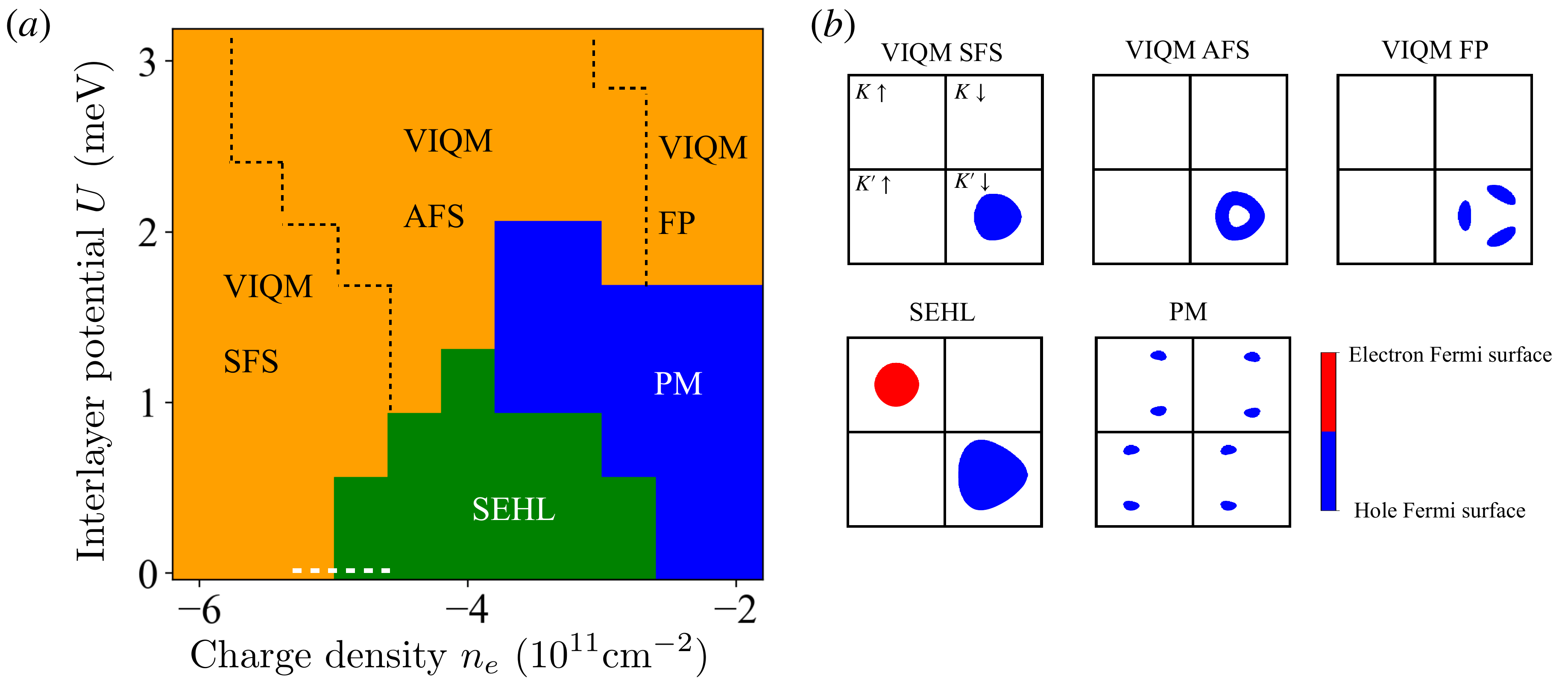}
    \caption{$(a)$ The mean-field phase diagram of rPG at hole doping side ($n_e<0$) near zero interlayer potential $(U)$ without SOC.  The energy crossing between VIQM and SEHL at $U=0$ along the white dashed line is shown in Fig.~\ref{fig:sup_2}.  $(b)$ The itinerant charge distributions across four spin-valley flavors $f\in\{K\uparrow,K\downarrow,K'\uparrow,K'\downarrow\}$ is shown for the competing groundstates in rPG at hole doping side.}
    \label{fig:sup_1}
\end{figure}
In the density $(n_e)$-interlayer potential $(U)$ phase space, three competing metallic phases emerge: the valley-Ising quarter metal (VIQM), the superpolarized electron-hole liquid (SEHL), and the $\mathcal{C}_3$ rotational symmetry broken unpolarized paramagnetic state (PM). These phases are distinguished by their relative population of itinerant charges at each spin-valley flavor, as described in the maintext. Fig.~\ref{fig:sup_1}.$(a)$ illustrates that at higher $U$, the groundstate is the VIQM phase, depicted in orange. As $n_e$ approaches neutrality, the VIQM phase undergoes two types of Lifshitz transitions, altering the topology of its Fermi surface. At lower $n_e$ and higher $U$, the Fermi surface of the VIQM phase consists of a single Fermi surface (SFS). With increasing charge density, the VIQM phase transitions through an annular Lifshitz transition, incorporating an electron pocket to form an annular Fermi surface (AFS). At even higher density, the AFS goes through a van-Hove singularity, transforming into three Fermi pockets (FP).

However, the groundstate near zero interlayer potential changes differently. As we increase the charge density near $U\sim0$, the VIQM phase with SFS undergoes a magnetic phase transition to the SEHL phase, depicted in green in Fig.\ref{fig:sup_1}.$(a)$. As the SEHL phase approaches neutrality, another magnetic phase transition occurs, leading to the emergence of the $\mathcal{C}_3$ symmetry broken unpolarized paramagnetic state (PM), shown in blue in Fig.\ref{fig:sup_1}.$(a)$. The Fermi surfaces across spin-valley flavors of all these competing states are depicted in Fig.\ref{fig:sup_1}.$(b)$. The red and blue colors in Fig.~\ref{fig:sup_1}.$(b)$ represent the electron and hole Fermi surfaces, respectively.

\begin{figure*}[h]
    \centering
    \includegraphics[width=0.5\columnwidth]{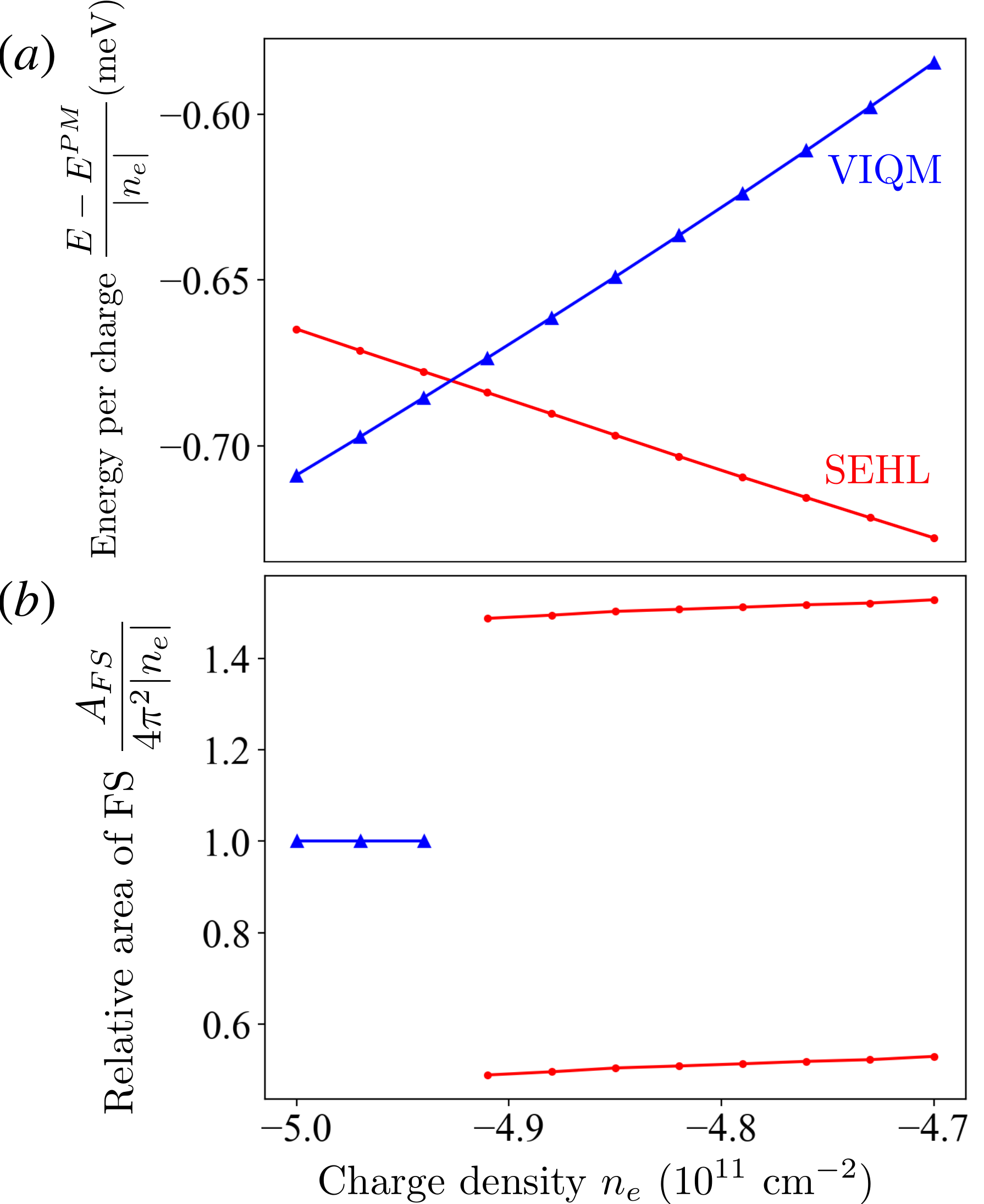}
    \caption{ $(a)$ The energy per charge of SEHL phase crosses the quarter metal (VIQM) near $n_e^*\sim-4.9\times 10^{11}$ cm $^{-2}$ at $U=0$ meV, and SEHL phase becomes groundstate at $n_e>n_e^*$. The energy per charge of each phase is measured with respect to paramagnetic phase at given charge density. $(b)$ The relative area(s) of groundstate Fermi surface(s) suggests the existence of two Fermi surface at $n_e>n_e^*$.}
    \label{fig:sup_2}
\end{figure*}

The Lifshitz transition of the VIQM phase is characterized by a first-order phase transition, primarily driven by long-range Coulomb repulsion \cite{PhysRevB.109.L060409}.
To discern the nature of the phase transition from VIQM to the SEHL phase, we assess the energy of each magnetic phase relative to the PM state at various charge densities, indicated by the horizontal dashed line in Fig.\ref{fig:sup_1}.$(a)$.

Fig.\ref{fig:sup_2}.$(a)$ illustrates the crossing of the SEHL energy per charge with the VIQM energy per charge near $n_e^*\sim -4.9\times 10^{11}$ cm $^{-2}$ at $U=0$, suggesting a first-order phase transition. Fig.\ref{fig:sup_2}.$(b)$ demonstrates the relative area of the groundstate Fermi surface(s), $\frac{A_{FS}}{4\pi^2|n_e|}$, at difference charge density at $U=0$. The experimental observation of two Fermi surfaces near $n_e\sim -5\times 10^{11}$ cm$^{-2}$ \cite{han2023orbital} further supports our findings. Note the phase boundary between SEHL and PM also exhibits a first-order characteristic.

\section{C: External field controlled coercive field}
One of the most important features of the SEHL phase is its exhibition of hysteresis in magnetic and electric fields. Furthermore, the significant magnetoelectric effect of the SEHL phase allows for the control of magnetic hysteresis through external electric field, and vice versa. In this section, we delve into how the magnetic coercive field $(B_c)$ varies with the electric field induced interlayer potential $(U)$, and conversely, how the electric coercive potential $(U_c)$ changes with the magnetic field $(B)$. Note, all the applied external fields are perpendicular to the material plane.
\begin{figure}[h]
    \centering
    \includegraphics[width=\columnwidth]{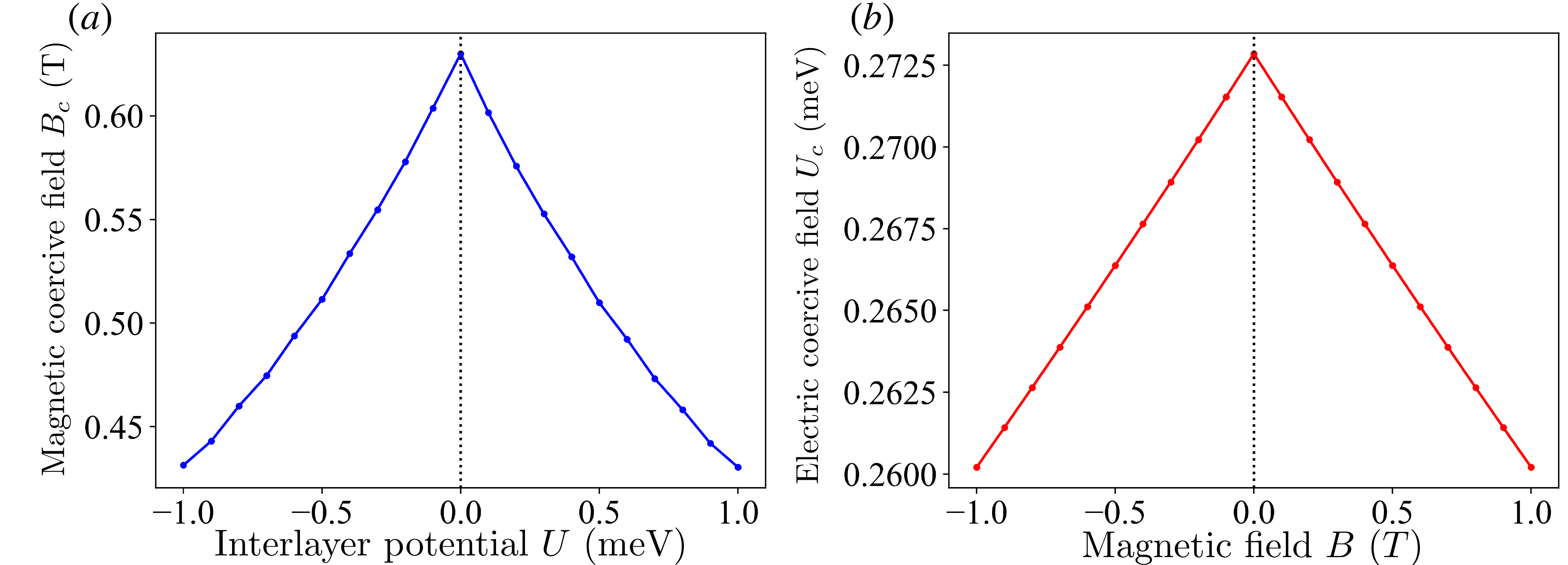}
    \caption{$(a)$ Magnetic coercive field $(B_c)$ decreases as $|U|$ increases. Similarly, $(b)$ electric coercive potential $(U_c)$ decreases as $|B|$ increases. }
    \label{fig:sup_4}
\end{figure}

The Ginzburg-Landau free-energy density $(F)$ used in our theory is expressed as:
\begin{align}
F = E - M^0_{s} B -  \frac{|n_e|}{2}\braket{\hat{\sigma}_z}^0  U - \alpha_{\text{orb}}  U B,
\end{align}
where  $E$ is energy density of the groundstate, $M_s^0$ denotes the spontaneous spin magnetization, $\braket{\hat{\sigma}_z}^0$ is the spontaneous layer polarization, and $\alpha_{\text{orb}}$ is the orbital magnetoelectric susceptibility. We use this above equation to determine the magnetic and electric coercive potentials.

The magnetic coercive field $(B_c)$ is determined by equating the difference in free energies between SEHL phases with opposite magnetization $(M)$ to the magnetic anisotropy energy $(E_{\text{aniso}})$,
\begin{align}
2(M_s^0 + \alpha_{\text{orb}}U)B_c = E_{\text{aniso}},
\end{align}
We estimate $(E_{\text{aniso}})$ as the energy difference between valley-Ising state and inter-valley coherent state, which in this case is around $0.2$ meV per charge. Here, $M_s^0$ denotes the spontaneous spin magnetization, and $\alpha_{\text{orb}}$ represents the orbital magnetoelectric susceptibility, also discussed in the maintext. The above equation yields the expression for the magnetic coercive field $B_c$:
\begin{align}
B_c = \frac{E_{\text{aniso}}}{2(M_s^0 + \alpha_{\text{orb}}U)}.
\end{align}
Fig.~\ref{fig:sup_4}.$(a)$ illustrates that $B_c$ is maximum at $U=0$. As the strength of $U$ increases on both sides, $B_c$ decreases. This behavior of $B_c$ across $U$ is also observed in experiment \cite{han2023orbital}, and the experimental estimate of $B_c$ at zero electric field aligns with our theoretical estimation at $U=0$.

A similar estimation of the electric coercive potential $U_c$ can be obtained by equating the free energy difference between two SEHL phases with opposite layer polarization $(\braket{\hat{\sigma}_z})$ and same magnetization $(M)$ to another magnetic anisotropy energy $E'_{\text{aniso}}$ as:
\begin{align}
2(|n_e|\braket{\hat{\sigma}_z}^0/2 + \alpha_{\text{orb}}B)U_c = E'_{\text{aniso}},
\end{align}
which leads us to the equation for the electric coercive potential $U_c$:
\begin{align}
U_c = \frac{E'_{\text{aniso}}}{2(|n_e|\braket{\hat{\sigma}_z}^0/2 + \alpha_{\text{orb}}B)}.
\end{align}
Here, we use $E'_{\text{aniso}}$ as the energy difference between the layer unpolarized paramagnetic (PM) phase and spontaneous layer polarized SEHL phase, and the value is around $1$ meV. The layer polarization is defined as $\braket{\hat{\sigma}_z}=\operatorname{Tr}(\hat{\sigma}_z\hat{\rho})/N$, where $\hat{\rho}$ is the density matrix, $\hat{\sigma}_z$ is the layer polarization operator, and $N=|n_e|A$ is total charge. Fig.~\ref{fig:sup_4}.$(b)$ indicates that $U_c$ is maximum at $B=0$, and decreases on either side with respect to the axis of the $B$ field.

A notable distinction between the magnetic and electric hysteresis of the SEHL phase is that the chirality of the magnetic hysteresis remains unaffected by the direction of the electric field, whereas the chirality of the electric hysteresis changes with alterations in the direction of the $B$-field. We have a detailed discussion about this behavior in the maintext.

In the following section, we investigate the origin of spontaneous $A_1-B_5$ layer polarization in the SEHL phase.

\section{D: layer polarization  of SEHL}
At low energy, the electronic bands in multilayer graphene reside on opposite sublattices of the outermost layers, namely $A_1-B_N$. Within the SEHL phase of pentalayer graphene $(N=5)$, two spin-valley flavors, which construct the Fermi surfaces, protect the Dirac touching between valance and conduction bands at $U=0$. Conversely, the remaining two flavors exhibit a band gap, and the sign of their mass is undetermined in absence of any external fields. We denote the sign of their mass by $0_\pm$ in the relative population ($\vec{\nu}$) description.

\begin{figure}[h]
    \centering
    \includegraphics[width=0.9\columnwidth]{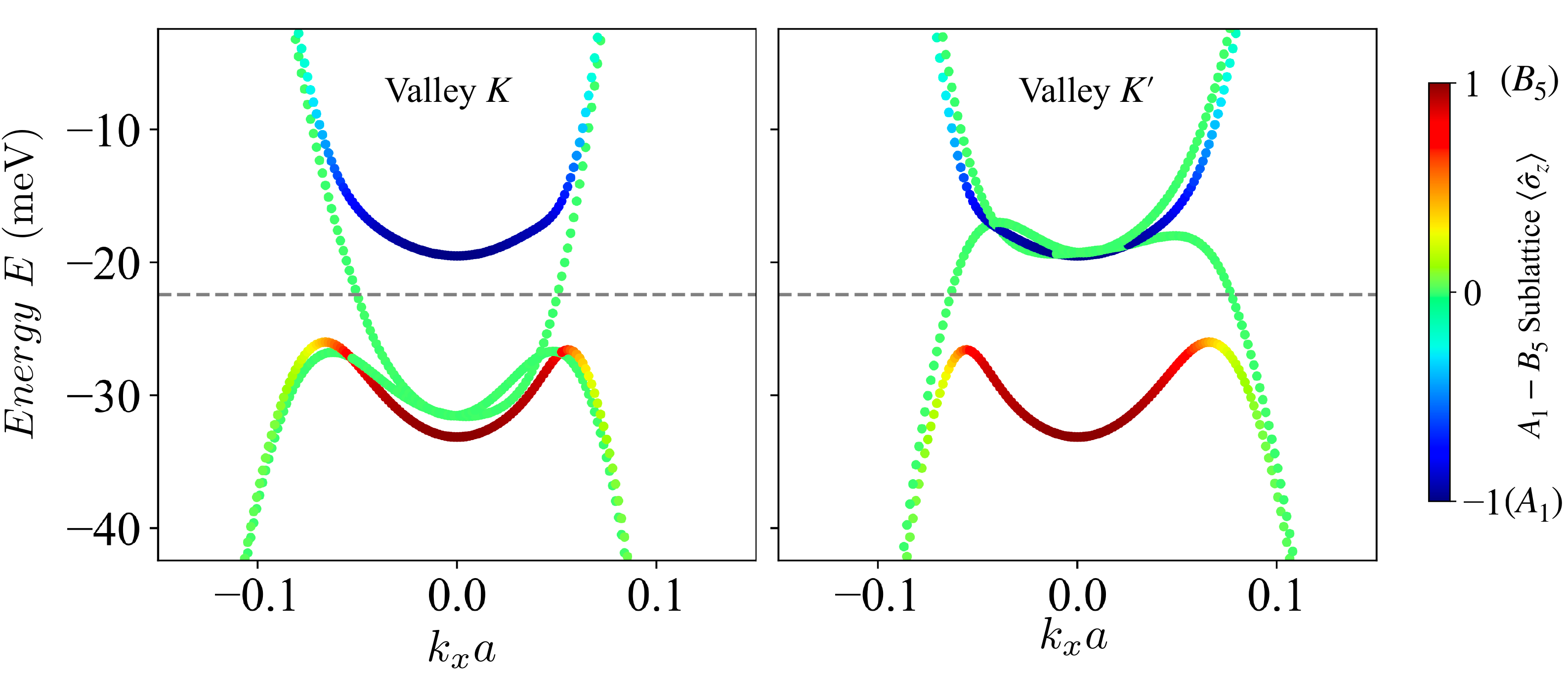}
    \caption{The layer polarization of SHEL phase at $U=0$ is shown along $k_y=0$. It shows the exchange interaction introduce spontaneous layer polarization in two of the spin-valley flavors.The relative charge distribution of the SEHL phase shown in this figure is $\vec{\nu}=(-f_e,0_-,0_-,f_h)$. }
    \label{fig:sup_3}
\end{figure}

The layer polarization operator in pentalayer graphene is defined as
\begin{align}
    \hat{\sigma}_z=\ket{B_5}\bra{B_5}-\ket{A_1}\bra{A_1},
\end{align}
and $\braket{\hat{\sigma}_z}_{n,\vec{k}}=\braket{\psi_{n,\vec{k}}|\hat{\sigma}_z|\psi_{n,\vec{k}}}=\pm 1$ indicates polarization in $B_5/A_1$ sublattice. 
Figure~\ref{fig:sup_3} indicates the band-projected layer polarization of SEHL phase $\vec{\nu}=(\nu_e,0_-,0_-,\nu_h)$ at $U=0$. The two metallic flavor bands, which form the Fermi surfaces of SEHL, maintain Dirac touchings between valence and conduction bands and do not polarize in any layer. This phase is spontaneously polarized in valley, spin, and layer $(B_5)$, but doesn't exhibit orbital magnetization $M_{\text{orb}}\propto\braket{\hat{\sigma}_z\hat{\tau}_z}$. Note, this phase is also favored by intrinsic SOC.

In the last section, we delve into the degeneracy of the SEHL phase in the absence of intrinsic spin-orbit coupling (SOC) and any external field.

\section{E: Degeneracy of SEHL phase without SOC}
In the absence of SOC and any external field, the hole Fermi surface of SEHL phase can be occupied by any one out of $4$ spin-valley flavors, and then the electron Fermi surface can occupy any one out of remaining $3$ flavors. Once the Fermi surfaces are occupied by two flavors, the remaining $2$ gapped flavors can introduce the mass terms in $2^2=4$ different ways. Hence, the total number of degenerate SEHL phases in absence of SOC and external fields are $n_d=4\times 3\times 2^2=48$. 

Depending on the flavors of the Fermi surfaces, the spin and valley polarization exceeds the saturation limit of magnitude $1$, which we mentioned as superpolarization in the maintext. Here we classify the degenerate SEHL phases into three distinct classes based on the magnitude of spin and valley polarization: $(a)$ If both spin and valley polarization exceeds saturated magnitude, we call them \textit{spin-valley superpolarization} (SVS). $(b)$ If  spin polarization doesn't exceed saturated level of magnitude, but magnitude of valley polarization$>1$, we call them \textit{valley superpolarization} (VS), and $(c)$ if magnitude of valley polarization remains $1$, while spin polarization exceeds magnitude of $1$, we classify them as \textit{spin superpolarization} (SS). Between  $16$ degenerate states in each class (SVS,VS,SS), $8$ of them are related to the time-reversal transformation to the other $8$ states. In the table bellow, we show spin,valley and layer polarization of these $8$ degeneracies from each class. We also indicate the order parameter $\braket{\hat{\sigma}_z\hat{\tau}_z}$ of these states, which is proportional to the spontaneous orbital magnetization$(M_{\text{orb}}^0)$. Note, the phases shown in red, are favored by intrinsic SOC without external field. Only four of the SVS phases are favoured by SOC and external fields, which participate in the electric and magnetic hysteresis.

\begin{table}[h]
  \begin{center}
  \vline
    \begin{tabular}{l|c|c|c|c|c|}
    \hline 
       & $
\vec{\nu}=\left(\frac{n_{K\uparrow}}{n_e},
\frac{n_{K\downarrow}}{n_e},
\frac{n_{K'\uparrow}}{n_e},
\frac{n_{K'\downarrow}}{n_e}\right)$& 
    $\braket{\hat{s}_z}$ & $\braket{\hat{\tau}_z}$ & $\braket{\hat{\sigma}_z}$ & $\braket{\hat{\sigma}_z\hat{\tau}_z}$
    \\
      \hline
       & \textcolor{red}{$(\nu_e,0_-,0_-,\nu_h)$} & \textcolor{red}{$\nu_h-\nu_e>1$} & \textcolor{red}{$\nu_h-\nu_e>1$}& \textcolor{red}{$+$}&\textcolor{red}{$0$}\\
      \cline{2-6}
       & $(\nu_e,0_-,0_+,\nu_h)$ & $\nu_h-\nu_e>1$ & $\nu_h-\nu_e>1$& $0$&$+$\\
      \cline{2-6}
       & $(\nu_e,0_+,0_-,\nu_h)$ & $\nu_h-\nu_e>1$ & $\nu_h-\nu_e>1$& $0$&$-$\\
       \cline{2-6}
       & $(\nu_e,0_+,0_+,\nu_h)$ & $\nu_h-\nu_e>1$ & $\nu_h-\nu_e>1$& $-$&$0$\\
       \cline{2-6}
       \text{SVS} & $(\nu_h,0_-,0_-,\nu_e)$ & $\nu_e-\nu_h<-1$ & $\nu_e-\nu_h<-1$& $+$&$0$\\
      \cline{2-6}
      & $(\nu_h,0_-,0_+,\nu_e)$ & $\nu_e-\nu_h<-1$ & $\nu_e-\nu_h<-1$& $0$&$+$\\
      \cline{2-6}
       & $(\nu_h,0_+,0_-,\nu_e)$ & $\nu_e-\nu_h<-1$ & $\nu_e-\nu_h<-1$& $0$&$-$\\
       \cline{2-6}
       & \textcolor{red}{$(\nu_h,0_+,0_+,\nu_e)$} &\textcolor{red}{ $\nu_e-\nu_h<-1$ }& \textcolor{red}{$\nu_e-\nu_h<-1$}& \textcolor{red}{$-$}&\textcolor{red}{$0$}\\
        \hline
        \hline
        & $(\nu_e,0_-,\nu_h,0_-)$ & $-1$ & $\nu_h-\nu_e>1$& $+$&$0$\\
      \cline{2-6}
       & \textcolor{red}{$(\nu_e,0_-,\nu_h,0_+)$} & \textcolor{red}{$-1$} & \textcolor{red}{$\nu_h-\nu_e>1$}& \textcolor{red}{$0$}&\textcolor{red}{$+$}\\
      \cline{2-6}
       & $(\nu_e,0_+,\nu_h,0_-)$ & $-1$ & $\nu_h-\nu_e>1$& $0$&$-$\\
       \cline{2-6}
       & $(\nu_e,0_+,\nu_h,0_+)$ & $-1$ & $\nu_h-\nu_e>1$& $-$&$0$\\
       \cline{2-6}
       \text{VS} & $(\nu_h,0_-,\nu_e,0_-)$ & $-1$ & $\nu_e-\nu_h<-1$& $+$&$0$\\
      \cline{2-6}
       & \textcolor{red}{$(\nu_h,0_-,\nu_e,0_+)$} &\textcolor{red}{ $-1$ }& \textcolor{red}{$\nu_e-\nu_h<-1$}& \textcolor{red}{$0$}&\textcolor{red}{$+$}\\
      \cline{2-6}
      & $(\nu_h,0_+,\nu_e,0_-)$ & $-1$ & $\nu_e-\nu_h<-1$& $0$&$-$\\
       \cline{2-6}
       & $(\nu_h,0_+,\nu_e,0_+)$ & $-1$ & $\nu_e-\nu_h<-1$& $-$&$0$\\
        \hline
        \hline
       & $(\nu_e,\nu_h,0_-,0_-)$ & $\nu_h-\nu_e>1$ & $-1$& $+$&$-$\\
      \cline{2-6}
      & \textcolor{red}{$(\nu_e,\nu_h,0_-,0_+)$} & \textcolor{red}{$\nu_h-\nu_e>1$} & \textcolor{red}{$-1$}& \textcolor{red}{$0$}&\textcolor{red}{$0$}\\
      \cline{2-6}
       & $(\nu_e,\nu_h,0_+,0_-)$ & $\nu_h-\nu_e>1$ & $-1$& $0$&$0$\\
       \cline{2-6}
       & $(\nu_e,\nu_h,0_+,0_+)$ & $\nu_h-\nu_e>1$ & $-1$& $-$&$+$\\
       \cline{2-6}
       \text{SS} & $(\nu_h,\nu_e,0_-,0_-)$ & $\nu_e-\nu_h<-1$ & $-1$& $-$&$+$\\
      \cline{2-6}
       & $(\nu_h,\nu_e,0_+,0_-)$ & $\nu_e-\nu_h<-1$ & $-1$& $0$&$0$\\
       \cline{2-6}
      & \textcolor{red}{$(\nu_h,\nu_e,0_-,0_+)$} & \textcolor{red}{$\nu_e-\nu_h<-1$} & \textcolor{red}{$-1$}& \textcolor{red}{$0$}&\textcolor{red}{$0$}\\
       \cline{2-6}
       & $(\nu_h,\nu_e,0_+,0_+)$ & $\nu_e-\nu_h<-1$ & $-1$& $+$&$-$\\
        \hline
    \end{tabular}
  \end{center}
  \end{table}

\end{document}